\documentclass[apj]{emulateapj}

\begin{document}

\title{Far-ultraviolet study of the local supershell GSH 006-15+7}

\author{Young-Soo Jo\altaffilmark{1,2}, Kyoung-Wook Min\altaffilmark{1}
, Kwang-Il Seon\altaffilmark{2,3}}

\email{email: stspeak@kasi.re.kr}

\altaffiltext{1}{Korea Advanced Institute of Science and Technology
(KAIST), 291 Daehak-ro, Yuseong-gu, Daejeon, Korea 305-701, Republic
of Korea}

\altaffiltext{2}{Korea Astronomy and Space Science Institute (KASI),
776 Daedeokdae-ro, Yuseong-gu, Daejeon, Korea 305-348, Republic of
Korea}

\altaffiltext{3}{Astronomy and Space Science Major, Korea University
of Science and Technology, 217 Gajeong-ro, Yuseong-gu, Daejeon,
Korea 305-350, Republic of Korea}

\begin{abstract}
We have analyzed the archival data of FUV observations for the
region of GSH 006-15+7, a large shell-like structure discovered by
\citet{mos12} from the \mbox{H\,{\sc i}} velocity maps. FUV emission
is seen to be enhanced in the lower supershell region. The FUV
emission is considered to come mainly from the scattering of
interstellar photons by dust grains. A corresponding Monte Carlo
simulation indicates that the distance to the supershell is 1300
$\pm$ 800 pc, which is similar to the previous estimation of 1500
$\pm$ 500 pc based on kinematic considerations. The spectrum at
lower Galactic latitudes of the supershell exhibits molecular
hydrogen fluorescence lines; a simulation model for this candidate
photodissociation region (PDR) yields an H$_2$ column density of
N(H$_2$) = 10$^{18.0-20.0}$ cm$^{-2}$ with a rather high total
hydrogen density of n$_H$ $\sim$ 30 cm$^{-3}$.
\end{abstract}


\section{Introduction}

Supershells are giant bubbles with a scale of hundreds of parsecs,
and are believed to be created by multiple stellar winds from OB
associations and supernova explosions \citep{hei79,hei84}.
Supershells play an important role in galaxy evolution since they
redistribute huge amounts of energy and material from the galaxy
disc to the galaxy halo. The interstellar matter swept up by
supershells forms cold and dense regions in which new star formation
is triggered. The typical lifetime of supershells is about 10$^7$
years \citep{ten88}, much longer than those of supernova remnants
\citep[SNRs;][]{fra94}. Supershells have been found with
observations of 21-cm \mbox{H\,{\sc i}} line emission originating
from the cold and dense shells
\citep{hei79,hei84,mcc02,ehl05,ehl13,sua14}, in which infrared
emission associated with dust is also observed. For younger
supershells, X-ray emission was  found from their inner hot gas
(T$>$10$^6$ K) \citep{hei99}. Furthermore, far ultraviolet (FUV)
emission lines such as \mbox{C\,{\sc iv}} and \mbox{O\,{\sc vi}}
were also observed from their interface region of T$\sim$10$^5$ K
between hot gas and cold ambient medium \citep{jo11}. Optical
emission lines such as H$\alpha$ were observed from their ionized
shells \citep{rey98,bou01}. Hence, multi-wavelength study for
supershells including FUV wavelengths may give us critical
information about their environment and evolution.

GSH 006-15+7 is one of the Milky Way supershells discovered recently
through a 21-cm line survey of the southern sky, which was compiled
as the Galactic All Sky Survey (\textit{GASS}) \citep{kal10,mos12}.
GSH 006-15+7 spans across the sky from \textit{l} = 356\degr to
\textit{l} = 16\degr in longitude and from \textit{b} = -28\degr to
\textit{b} = 2\degr in latitude, which is translated into the
physical size of 790 $\times$ 520 pc at a distance of 1500 $\pm$ 500
pc. \citet{mos12} also discussed in detail the physical properties
of the supershell. The structure of the supershell is most obvious
in the range of v$_{LSR}$$\sim$5 km s$^{-1}$ to v$_{LSR}$$\sim$11 km
s$^{-1}$, with a systemic velocity of $\sim$7 km s$^{-1}$, whereas
it is difficult to identify the shell-like structure in the maps of
the lower velocity channels of v$_{LSR}$ $<$ 5 km s$^{-1}$, probably
because of the contamination by local emission. Its dynamical age
has been estimated to be $\sim$15 $\pm$ 5 Myr, which is consistent
with the ages found for similar Galactic supershells with
well-defined shapes. The formation energy was estimated to be
$\sim$10$^{52}$ ergs, with a mass of $\sim$3 $\pm$ 2 $\times$ 10$^6$
M$_\sun$ and an expansion velocity of 8 km s$^{-1}$. It was
suggested that the energy source of the expansion of GSH 006-15+7 is
likely to be associated with Sgr OB association 1 and some open
clusters nearby. However, the absence of high-mass stars in these
stellar groups, which are able to produce significant stellar winds,
indicates the expansion was at least partially made by supernova
explosions and the winds of progenitor stars. \citet{mos12} found
some evidence for the breakout of the supershell at (\textit{l},
\textit{b}) = $\sim$(10\degr, -25\degr) based on the infrared and
\mbox{H\,{\sc i}} maps, indicating mass and energy transfer from the
Galactic disk to the halo.

Whereas the physical characteristics described above were obtained
from multi-wavelength data, FUV observations have not been reported
yet. It is well known that interstellar FUV emission is produced in
the regions of a cooling hot gas \citep{seo06,kim07}, from
scattering by dust grains \citep{jo12,lim13,cho13}, and in
photo-dissociation regions (PDRs) \citep{ryu06,lee06}; of these,
dust scattered FUV emission is regarded to be the dominant component
\citep{seo11}.

We present in this paper the results of analysis based on archival
data of FUV observations. Section 2 describes observational results
of the FUV continuum image and spectrum for the region of the
supershell. In Section 3, a distance to the supershell is estimated
based on three-dimensional multi-scattering simulations of dust
scattering. The results of Monte Carlo simulations for the PDR
region of the supershell are also presented in Section 3. A summary
of the present study is given in Section 4.

\begin{figure*}
 \begin{center}
  \includegraphics[height=8.0cm]{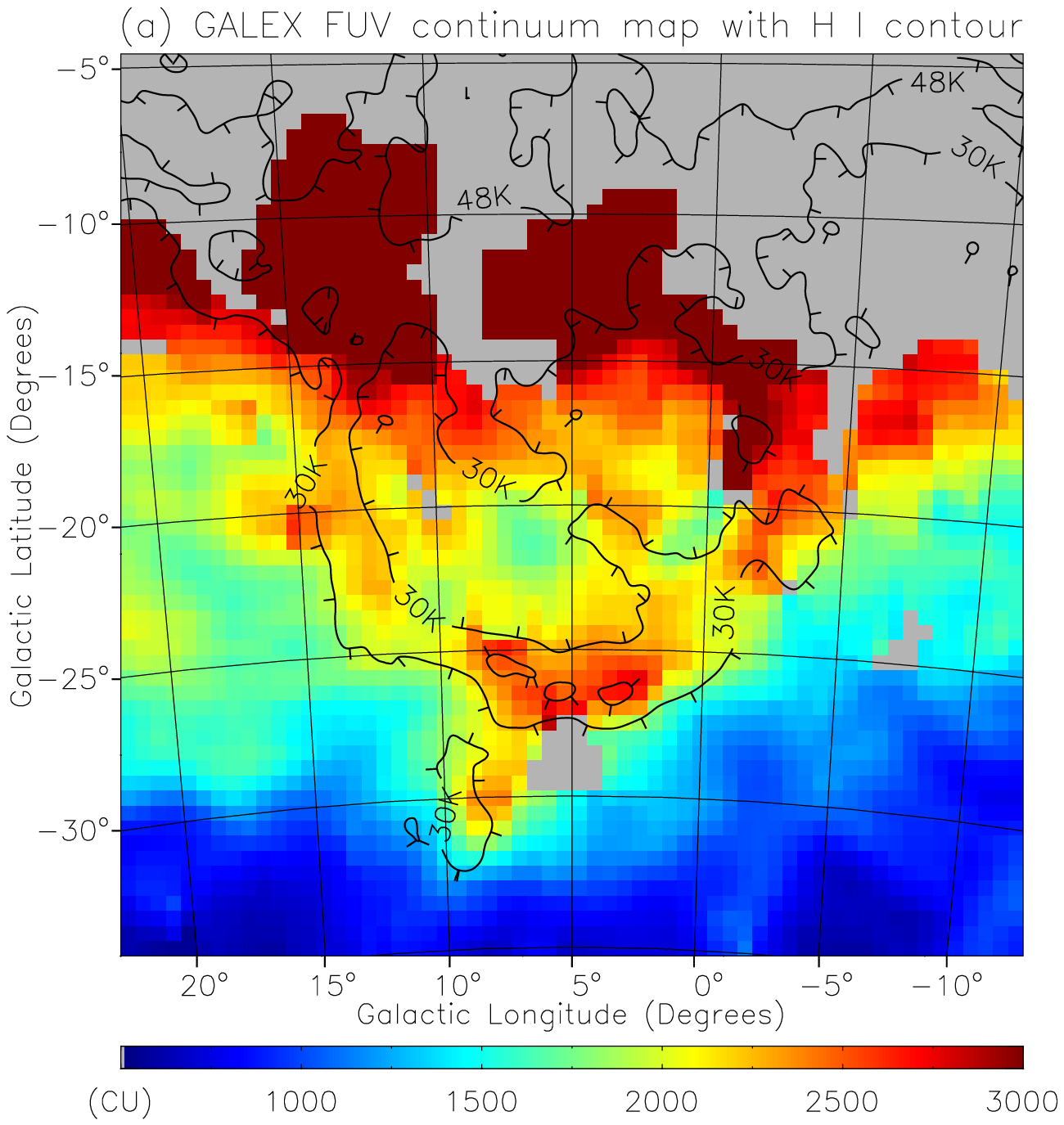}\hspace{10pt}
  \includegraphics[height=8.0cm]{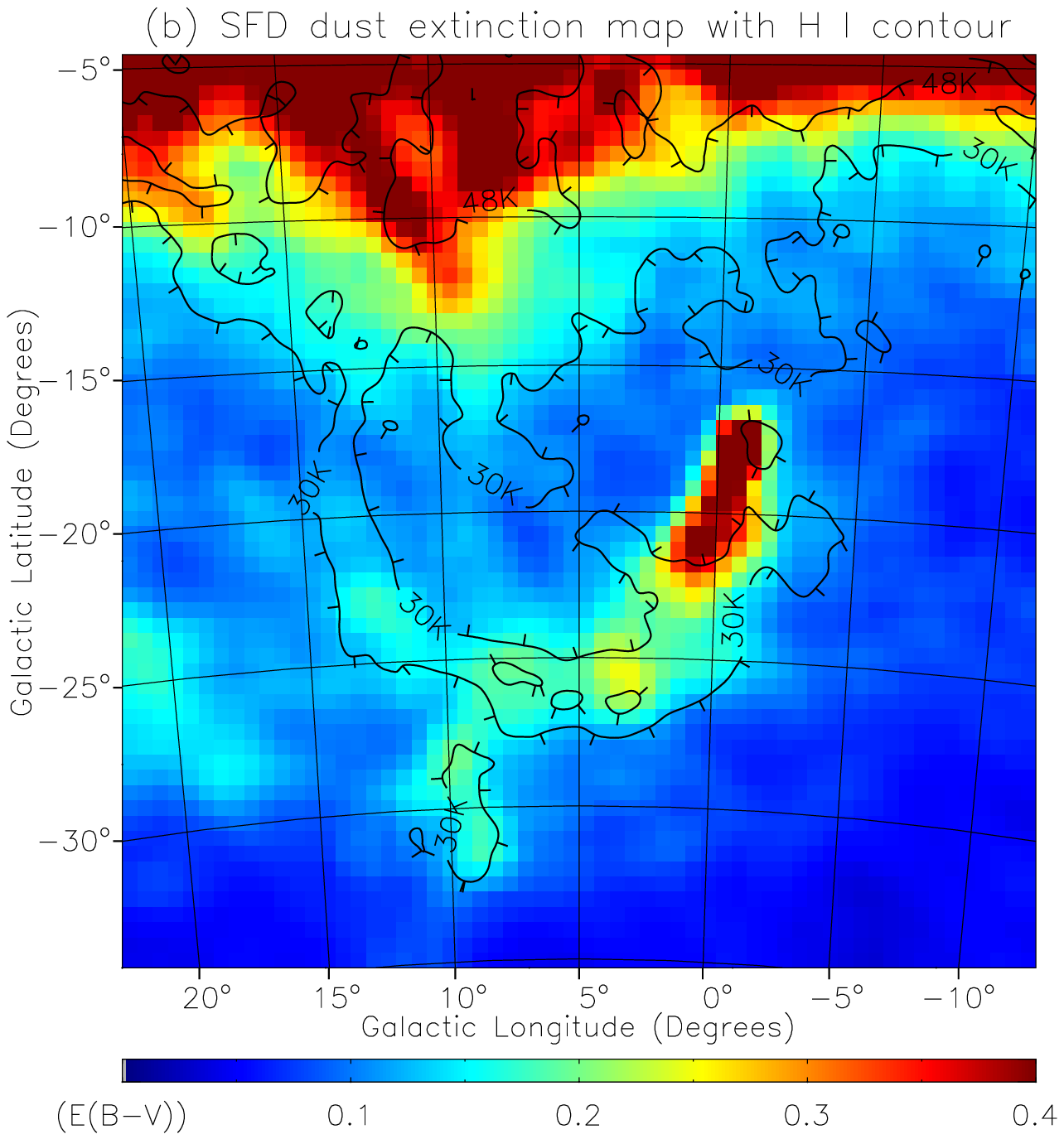}\\
  \vspace{10pt}
  \includegraphics[height=8.0cm]{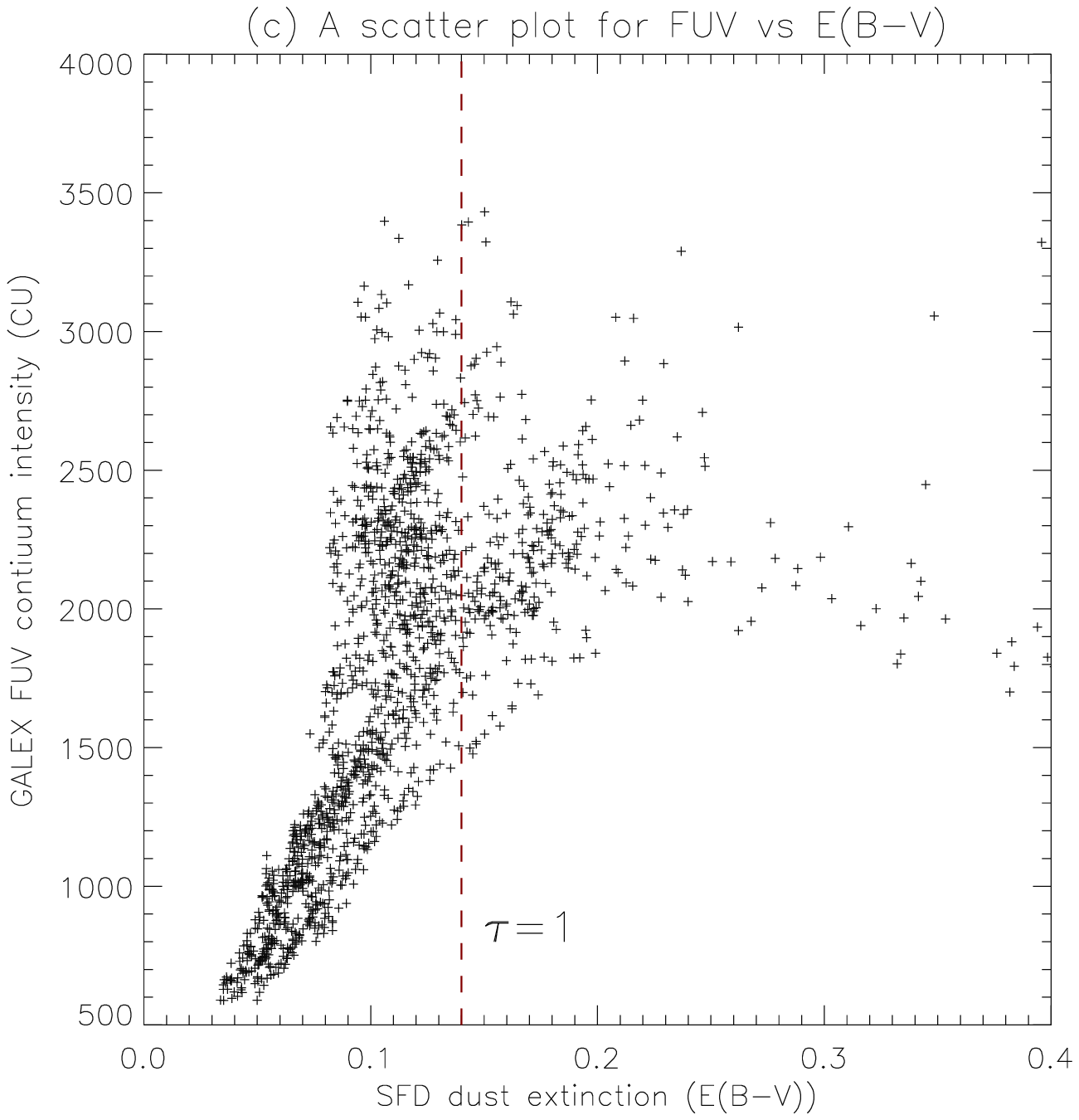}\\
 \end{center}
\caption{(a) \textit{GALEX} FUV continuum map, (b) SFD dust
extinction map, and (c) a scatter plot for the FUV intensity against
dust extinction.  The gray color in (a) indicates the missing data
points of the \textit{GALEX} observations. FUV intensity in (a) is
given in Continuum Units (photons s$^{-1}$ cm$^{-2}$ sr$^{-1}$
{\AA}$^{-1}$) and dust extinction in (b) is expressed in
\textit{E(B-V)}. The scatter plot of (c) was obtained from a
pixel-by-pixel comparison of (a) and (b) for the lower supershell
region. \label{fig:fuv}}
\end{figure*}


\section{Observations and Data Reduction}

The present study primarily used the FUV observations on the
rectangular region spanning from (\textit{l}, \textit{b}) =
(-10\degr, -35\degr) to (20\degr, -5\degr) which included part of
the supershell GSH006-15+7. A study of continuum emission was
conducted with the data set obtained by the Galaxy Evolution
Explorer \citep[\textit{GALEX};][]{mor07}. \textit{GALEX} carried
out an ultraviolet all sky survey in the FUV (1350--1750 {\AA})
band, but it does not provide spectral information. Hence, a
spectral study was proceeded with the observations from the
Far-ultraviolet IMaging Spectrograph (\textit{FIMS}, also known as
Spectroscopy of Plasma Evolution from Astrophysical Radiation or
\textit{SPEAR}) on board the Korean microsatellite STSAT-1.

\textit{FIMS} contains dual FUV imaging spectrographs, referred to
as the S-band (900--1150 {\AA}) and L-band (1350--1750 {\AA}).
\textit{FIMS} is the main payload of STSAT-1, which was operated for
a year and a half after its launch on September 27, 2003, in a 700
km sun-synchronous orbit. \textit{FIMS} was designed to have a large
field of view (7\fdg5 $\times$4 \farcm3 in L-band) with angular
resolution of $\sim$5\arcmin, which is suitable for observations of
diffuse targets, while its spectral resolution is rather moderate
($\lambda$/$\Delta\lambda$ $\sim$ 550). More information on the
\textit{FIMS} mission and instruments can be found in
\citet{ede06a,ede06b}. We also used the dust extinction map based on
the \citeauthor*{sch98} \citep[SFD;][]{sch98} for comparisons with
FUV observations.

\begin{figure*}
 \begin{center}
  \includegraphics[width=15cm]{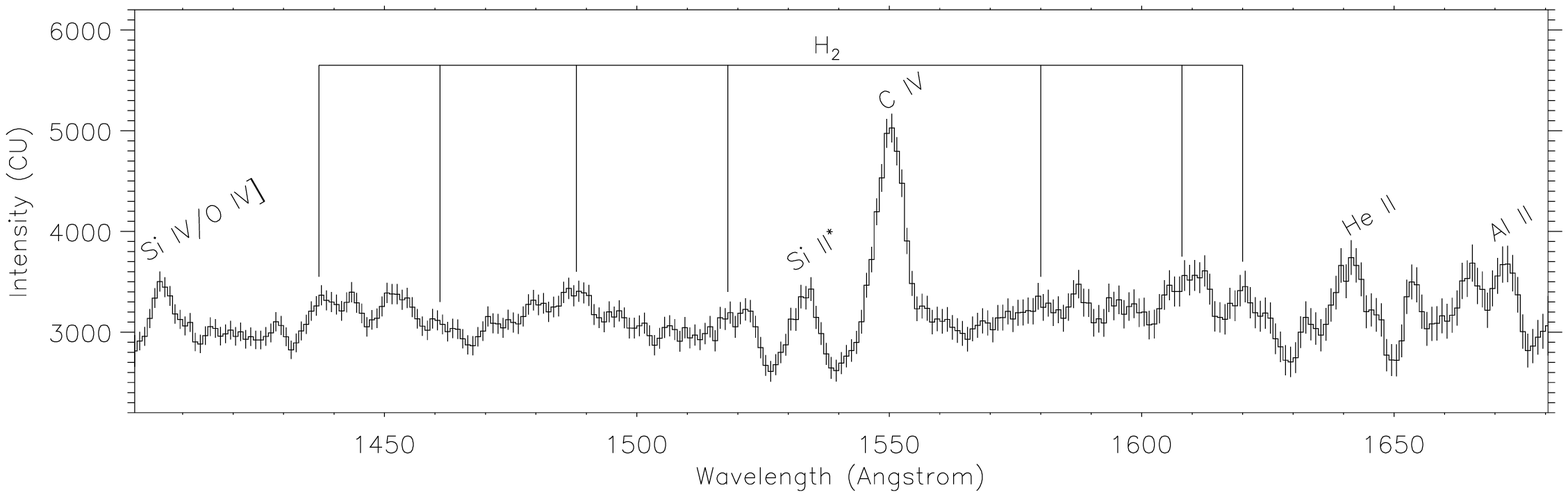}\\
 \end{center}
\caption{\textit{FIMS} FUV spectrum of GSH 006-15+7 over the
rectangular region spanning from (\textit{l}, \textit{b}) =
(-10\degr, -35\degr) to (20\degr, -5\degr), binned with 1 {\AA} and
smoothed with 3 {\AA} \label{fig:spec}}
\end{figure*}

Figure \ref{fig:fuv} shows the (a) \textit{GALEX} FUV continuum map,
(b) SFD dust extinction map over the rectangular region spanning
from (\textit{l}, \textit{b}) = (-10\degr, -35\degr) to (20\degr,
-5\degr), and (c) a scatter plot for the FUV intensity against dust
extinction for the region spanning from (\textit{l}, \textit{b}) =
(-5\degr, -35\degr) to (15\degr, -15\degr), which is referred to as
the lower supershell region henceforth. \mbox{H\,{\sc i}} contours
at the systemic velocity of 7 km s$^{-1}$, following the result of
\citet{mos12}, are overplotted in Figures \ref{fig:fuv}(a) and
\ref{fig:fuv}(b) \citep[The Parkes Galactic All-Sky Survey
(\textit{GASS}) dataset;][]{mcc09,kal10}. The Figure
\ref{fig:fuv}(a) and \ref{fig:fuv}(b) were plotted with 0\fdg5 pixel
size and smoothing of the image to have 1\degr resolution. The
\textit{GALEX} observations avoided bright stars intentionally,
leaving some regions unobserved. Most of these missing areas were
filled for the region of \textit{b} $<$ -15\degr, where the lower
supershell region resides, with coarse-grained binning and
smoothing, but some regions still remain unfilled, especially in the
region of the Galactic disk, as Figure \ref{fig:fuv}(a) shows. The
dust extinction map of Figure \ref{fig:fuv}(b) shows extremely high
optical depth in the region of \textit{b} $>$ -15\degr near the
Galactic disk, probably due to the foreground dust. The shell-like
feature associated with GSH 006-15+7 is not seen in this optically
thick region. Hence, we focus only on the lower region of \textit{b}
$<$ -15\degr in the present analysis. The scatter plot of Figure
\ref{fig:fuv}(c) was constructed from a pixel-by-pixel comparison
between Figure \ref{fig:fuv}(a) and \ref{fig:fuv}(b) for the lower
supershell region.

What is notable in the FUV continuum map of Figure \ref{fig:fuv}(a)
is that the intensity is higher at the lower supershell region of
\textit{b} $<$ -15\degr than its surroundings. This is considered to
result from dust scattering of the region since the lower supershell
region has higher extinction levels than its neighbors, as shown in
Figure \ref{fig:fuv}(b). We note that the optical depth for the
\textit{E(B-V)} value of 0.14 corresponds to $\tau$ $\sim$ 1 at
$\lambda$ = 1565 {\AA}, which is indicated by the vertical dashed
line in Figure \ref{fig:fuv}(c).

A strong correlation between the FUV intensity and \textit{E(B-V)}
is seen in Figure \ref{fig:fuv}(c) for the optically thin region of
$\tau$ $<$ 1, except the vertical branch above the FUV intensity of
1500 CU. This agrees well with the general notion that FUV continuum
intensity correlates with \textit{E(B-V)} in an optically thin
region \citep{hur91,seo11}. The vertical branch, where data points
are scattered up to $\sim$ 3500 CU, are the bright pixels
corresponding to the regions of \textit{b} $>$ -18\degr.
 Being close to the Galactic plane, they represent the
scattered component of the high-intensity radiation fields from the
Galactic plane. We also note that FUV intensity is anti-correlated
with dust extinction for \textit{E(B-V)} above 0.2, which is
reasonable for an optically thick region such as R CrA of the
present case, as can be seen in Figure \ref{fig:fuv}(b). The dense
cloud at (\textit{l}, \textit{b}) $\sim$ (0\degr, -18\degr) in
Figure \ref{fig:fuv}(b), corresponding to the partial extinction
seen in the FUV continuum map of Figure \ref{fig:fuv}(a), is the R
CrA molecular cloud \citep{ros78} which is located at a distance of
$\sim$170 pc from the Sun \citep{knu98}.

Figure \ref{fig:spec} shows the spectrum made with \textit{FIMS}
over the rectangular region spanning from (\textit{l}, \textit{b}) =
(-10\degr, -35\degr) to (20\degr, -5\degr) of GSH 006-15+7. Bright
point sources in the region were removed from the data set when
obtaining the above spectrum. More specifically, the pixels
corresponding to the bright stars in the TD-1 catalog with 1565
{\AA} band flux of above 10$^{-11}$ erg {\AA}$^{-1}$ s$^{-1}$
cm$^{-2}$ were removed, together with those within a radius of
2\degr around the stars whose intensities were more than twice as
high as the median values. Among various emission peaks, the
\mbox{C\,{\sc iv}} emission line is readily noticeable in the
spectrum, together with H$_2$ fluorescent emission features. While
the molecular hydrogen emission is associated with the shell
boundaries, it will be shown that the \mbox{C\,{\sc iv}} emission
does not seem to be related to the shell.

\begin{figure*}
 \begin{center}
  \includegraphics[height=8.0cm]{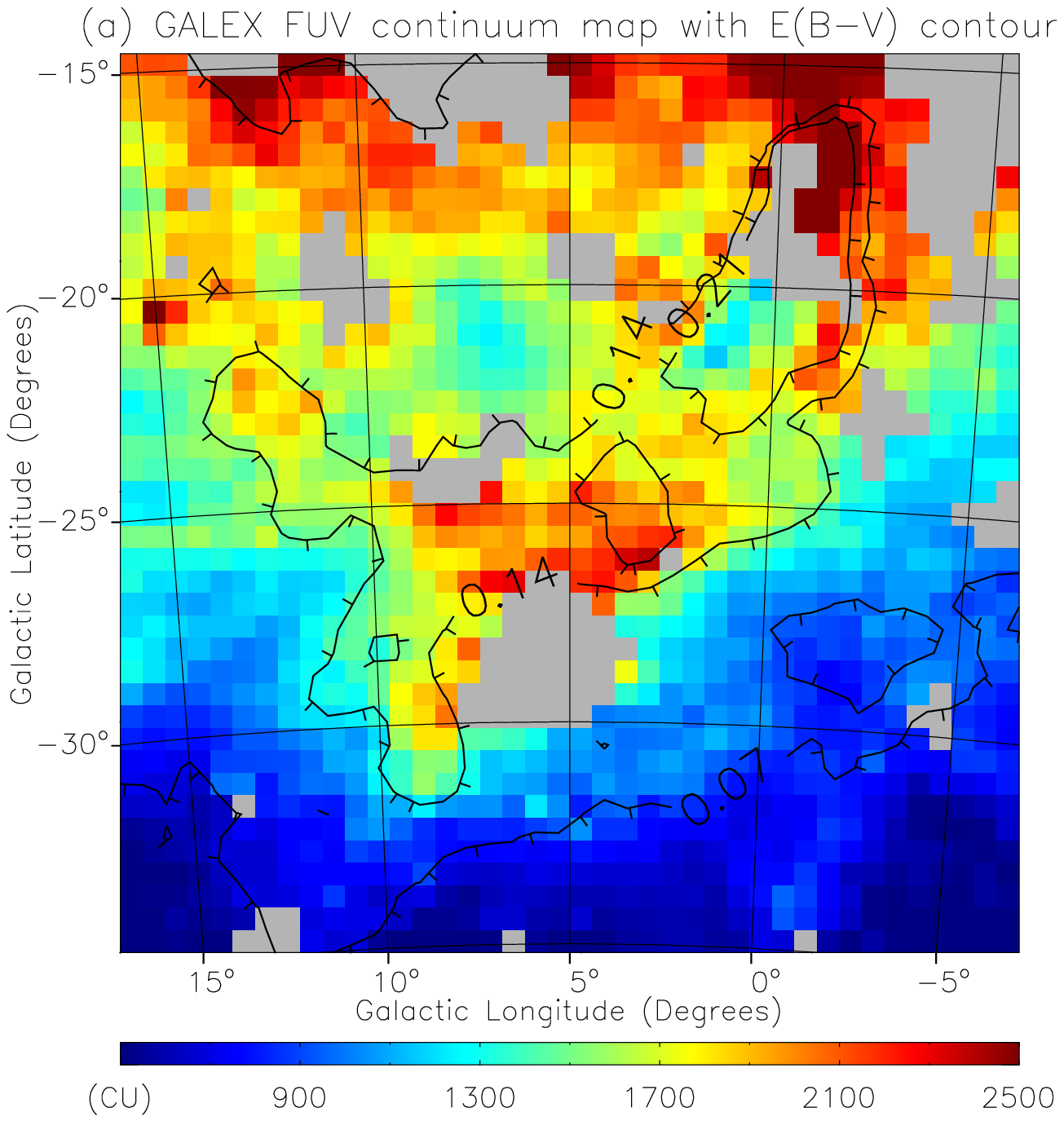}\hspace{10pt}
  \includegraphics[height=8.0cm]{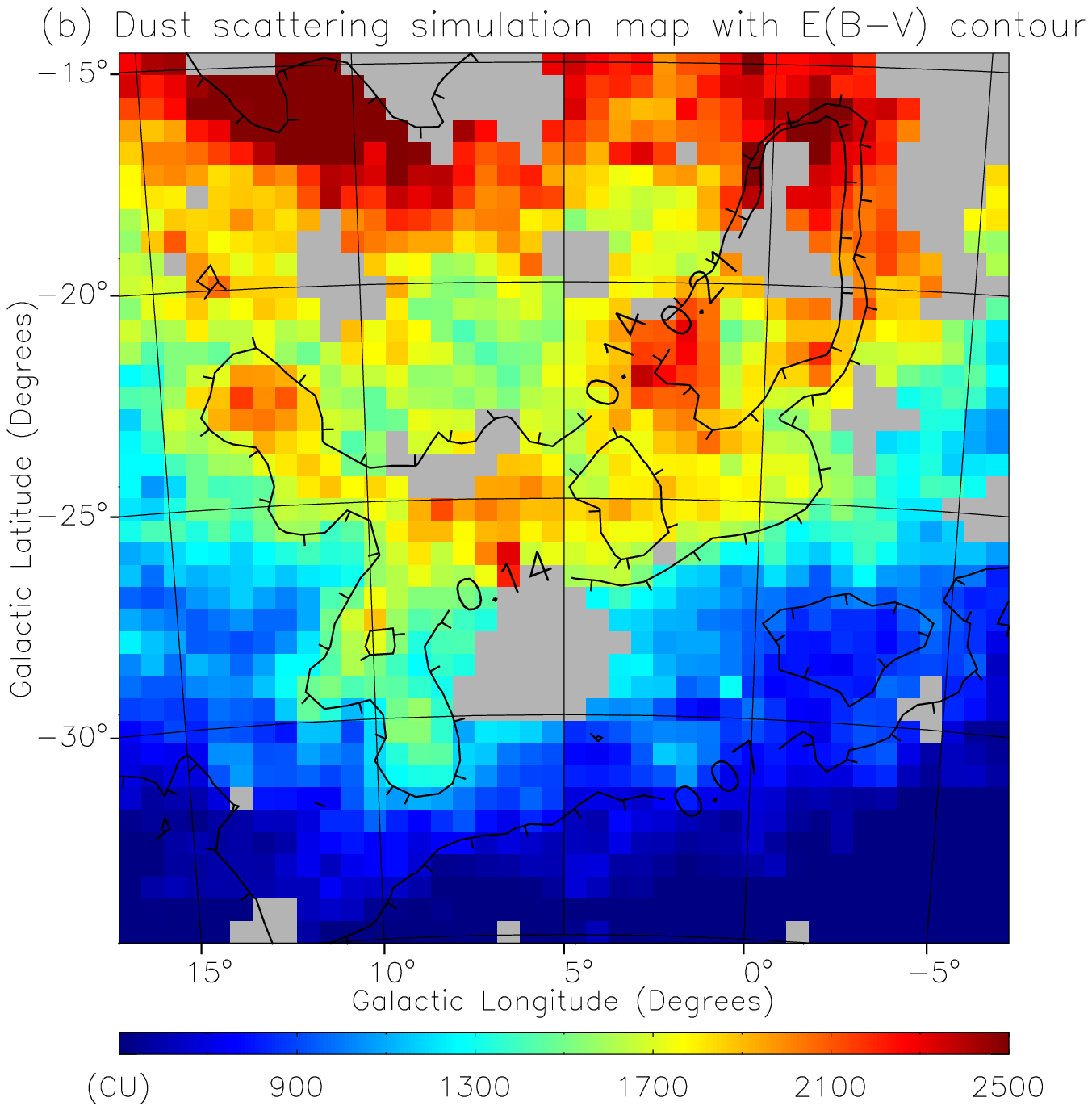}\\
  \vspace{10pt}
  \includegraphics[height=8.0cm]{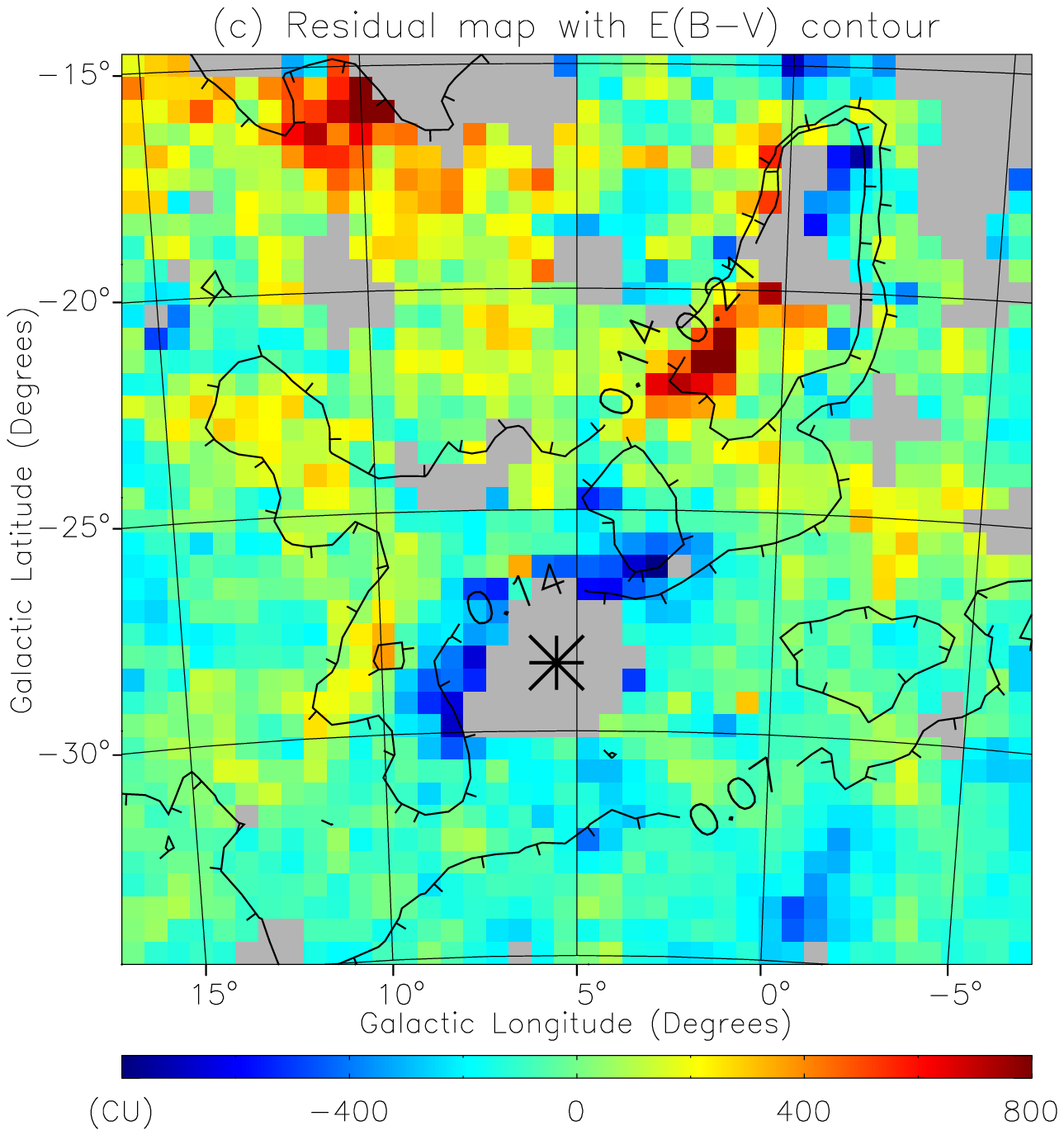}\hspace{10pt}
  \includegraphics[height=7.8cm]{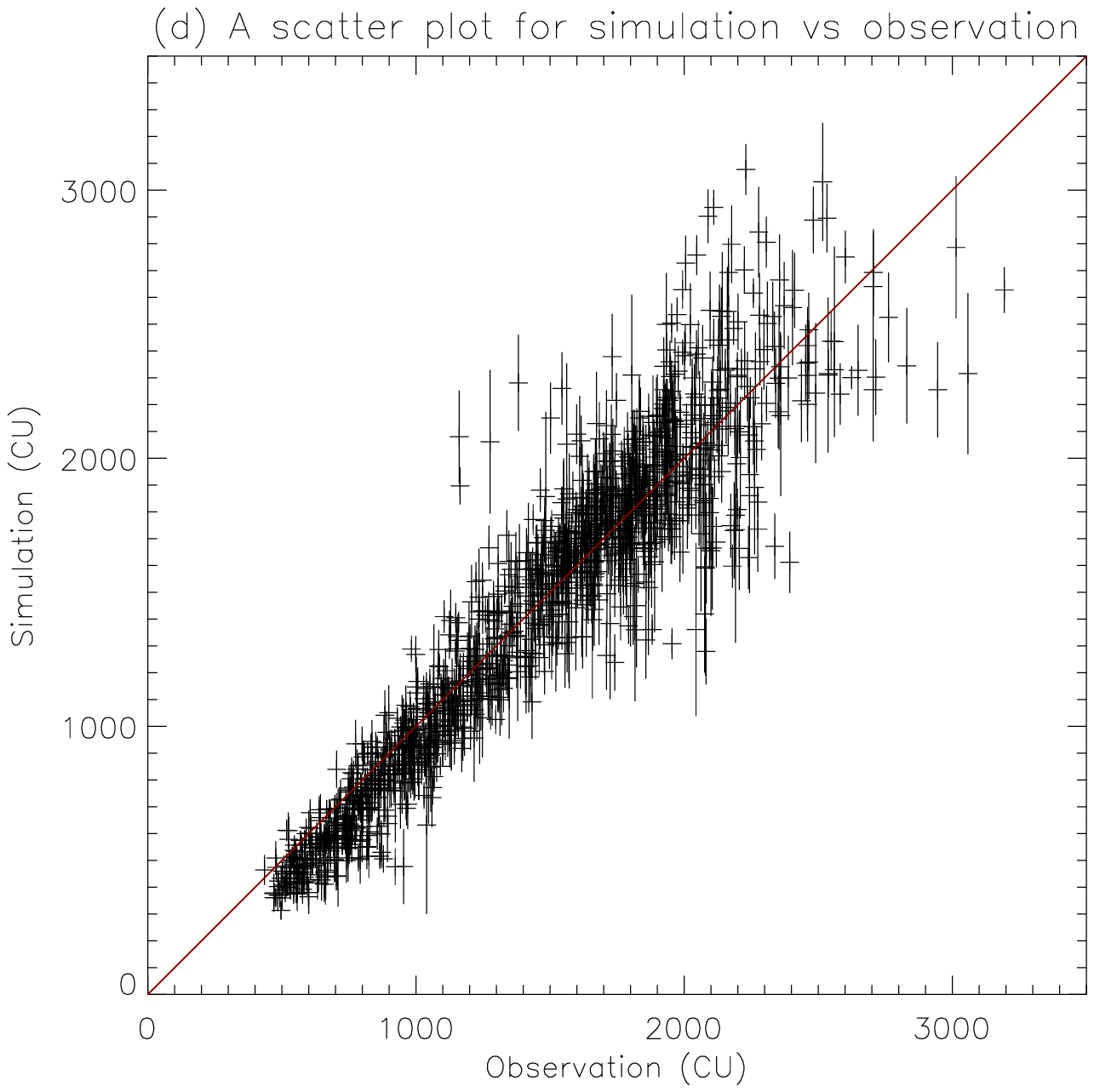}\\
 \end{center}
\caption{(a) Observed FUV continuum map with the contributions other
than dust scattering removed, (b) best-fit dust scattering
simulation result, (c) residuals of the simulation result subtracted
by the observation, and (d) a pixel-to-pixel scatter plot of the
simulation map against the observation map. In (a), (b), and (c),
dust contours are overplotted and the continuum intensity is given
in CU. The gray color in (a), (b), and (c) corresponds to the
missing data points of the GALEX observations. The asterisk symbol
in (c) indicates the UV bright star HD189103. \label{fig:dust}}
\end{figure*}


\section{Modeling and Discussion}

\subsection{Dust scattering simulation}

As mentioned previously, the FUV continuum emission associated with
GSH 006-15+7 is mainly from dust scattering. Hence, we attempted
simulations of dust scattering to obtain the distance to the
supershell region, by comparing the observation with a simulation
for the FUV continuum, as was done previously for many individual
targets \citep{jo12,lim13,cho13}. This will produce an independent
measurement of the distance that we will compare with the 1500 $\pm$
500 pc distance estimate that \citet{mos12} calculated from the
\mbox{H\,{\sc i}} gas kinematics. For comparison with the dust
scattering simulation, the observed FUV map should also have only
the contribution from dust scattering. We found that other sources
of FUV emission have only a small effect on this region: ion line
emission and H$_2$ fluorescent emission are $\sim$18 percent of the
total FUV intensity, and the two photon continuum emission ranges
from $\sim$20 to $\sim$90 CU, based on the Southern H-Alpha Sky
Survey Atlas \citep{gau01}. We used the conversion factor of 28.4 CU
for 1 Rayleigh of H$\alpha$ intensity to obtain the two photon
continuum emission \citep{seo11}. These contributions were removed
from the map of Figure \ref{fig:fuv}(a) and the result is shown in
Figure \ref{fig:dust}(a) with dust extinction contours at
\textit{E(B-V)} of 0.07, 0.14 and 0.21 for the lower supershell
region.

The simulation code employs a Heney-Greenstein function for the
scattering model: it describes light scattering off grains using the
albedo and the phase function asymmetry factor, or the g-factor, as
free parameters \citep{hen41}. In the present simulations, however,
the albedo and g-factor are fixed at 0.4 and 0.6, respectively,
based on the theoretical estimations for carbonaceous-silicate
grains \citep{dra03}, and Monte Carlo simulations are performed for
various distributions of dust to obtain the best dust distribution
model by comparing the simulated FUV intensity with the observed
one. We take a rectangular box of 280 $\times$ 280 $\times$ 700
cells as the simulation domain, corresponding to 1400 pc $\times$
1400 pc $\times$ 3500 pc with the longest dimension along the line
of sight. The center of the map of (\textit{l}, \textit{b}) =
(5\degr, -25\degr) corresponds to the sightline of the central axis,
and the Sun is located in the center of the front face. The 1400 pc
$\times$ 1400 pc size at the rear face spans an angular size of
22\fdg6 $\times$ 22\fdg6 on the sky, encompassing the lower
supershell region of 20\degr $\times$ 20\degr.

The dust distribution of the supershell is idealized to be a single
slab whose distance and thickness are to be determined in comparison
with the observed FUV map. We note that the region also has a local
cloud R CrA, as well as a possible contribution from the foreground,
both of which must be subtracted from the SFD map to obtain the
effect of dust solely associated with the supershell.

The foreground contribution is modeled as follows. We divide the SFD
map of \textit{l} = -5\degr to 15\degr into 500 vertical strips and
fit each strip with a linear function that generally follows the
minimum of \textit{E(B-V)}. Each linear fit, given as a function of
latitude, represents the \textit{E(B-V)} value integrated along the
sightline of the given latitude. We distribute dust along the
sightline according to the distance from the Galactic plane assuming
that dust distribution has an exponential decrease outwards from the
Galactic plane with a scale height of 100 pc in view of the previous
studies \citep{par45,mis06}. With this distribution of dust, 88
percent of the total foreground dust is located within 500 pc from
the Sun, and 99 percent within 1000 pc. The amount of dust
associated with the supershell is determined by subtracting these
foreground values from those of the SFD map at corresponding
locations. For the region of the supershell co-located with R CrA,
\textit{E(B-V)} = 0.1, the mean value of the nearby supershell
region is assigned to the supershell. The dust layer corresponding
to R CrA is placed at 170 pc from the Sun with a thickness of 5 pc.
With this given distribution of dust for the foreground and R CrA,
the distance to the dust clouds corresponding to the supershell is
decided by comparing the simulation with the observation for the FUV
continuum.

The source photons to be scattered by dust are difficult to model
for distant targets such as GSH 006-15+7 since stars have not been
completely surveyed to comparable distances. In fact, the number of
observed stars decreases rapidly when the distance from the Sun is
greater than $\sim$500 pc. Hence, we consider two types of source
photons. First, the photons from the bright stars of the TD-1
catalogs are scattered by the foreground dust and the R CrA cloud.
Second, for scattering from the supershell GSH 006-15+7, we adopted
the interstellar radiation field (ISRF) obtained from the {\sc
GALPROP} code \citep[http://galprop.stanford.edu/ and ][]{mos06}.
{\sc GALPROP} provides energy density at a specified wavelength, for
which we adopt $\lambda$ = 1565 {\AA}, as a function of location. It
should be noted that $\lambda$ = 1565 {\AA} is one of the passbands
of the TD-1 catalogs and is located nearly at the center of the FUV
passband of \textit{GALEX}. Furthermore, we confirmed that {\sc
GALPROP} gives a specific energy density at $\lambda$ = 1565 {\AA},
which is very similar to the mean energy density averaged over the
wavelength range of 1350--1750 {\AA}. We assume the photons,
corresponding to the specified energy density at a given location,
originate from the Galactic plane and propagate vertically to the
supershell, which we believe is reasonable as the supershell is
located in the high latitude region and there are not many nearby
stars. The scattered photons from the supershell are allowed to
interact with the foreground dust as well as R CrA.

With these photon sources, together with the model distribution of
foreground dust and R CrA, simulations were conducted for different
sets of distance and thickness of the supershell. The distance is
varied from 500 pc to 2500 pc with 100 pc steps, and the thickness
from 20 pc to 200 pc with 20 pc steps, and the resulting images are
compared with the observation for the region below \textit{b} =
-18\degr and excluding R CrA to obtain the best-fit parameters using
the chi-squared minimization method. We note that the region close
to the Galactic plane above \textit{b} = -18\degr is excluded
because of the contamination by the foreground local dust. Figure
\ref{fig:dust}(b) shows the simulated FUV map with 0\fdg5 $\times$
0\fdg5 pixel bins for the best-fit parameters: 1300 pc for the
distance to the center of the dust slab and 60 pc for the thickness
of the slab. The 1-sigma confidence range of the fit is 1300 $\pm$
800 pc for the distance to the center of the dust slab, which is in
good agreement with the estimation of 1500 $\pm$ 500 pc by
\citet{mos12}. We have also varied albedo and g-factor over a
reasonable range, with albedo from 0.3 to 0.5 and g-factor from 0.4
to 0.7 based on \citet{dra03}, respectively, to check the dependence
of the distance estimation on the changes of these optical
parameters. We obtained the distance in the range of 800--2000 pc,
which is within the 1-sigma range of the distance obtained with
albedo = 0.4 and g-factor = 0.6. While the range of thickness is not
well constrained in the present simulations, the best-fit value of
60 pc is similar to the $\sim$53 pc estimated from the column
density of N$_H$ $\sim$ (2.7--7.7) $\times$ 10$^{20}$ cm$^{-2}$ and
the number density of n$_H$ $\sim$ (1.7--4.8) atoms cm$^{-3}$ in
\citet{mos12}.

Figure \ref{fig:dust}(c) is a residual map for the best-fit
parameters that shows the difference between the simulation result
of Figure \ref{fig:dust}(b) and the observation of Figure
\ref{fig:dust}(a). Figure \ref{fig:dust}(d) is a scatter plot
obtained from a pixel-to-pixel comparison between the observation
and the simulation maps. While the simulation and the observation
are in general agreement over the entire region, we see that there
are several local regions with marked differences. For example, the
simulated intensity is much higher than the observed intensity in
the region around (\textit{l}, \textit{b}) = (10\degr, -16\degr) in
Figure \ref{fig:dust}(c), which is located close to the Galactic
plane and affected by the foreground local dust. The large
fluctuations above 2000 CU of the observed intensity in Figure
\ref{fig:dust}(d) correspond to this region. Another excess in
simulated intensity is seen near (\textit{l}, \textit{b}) = (2\degr,
-21\degr), which is close to R CrA, implying that the discrepancy is
due to the imperfect removal of the local dust effect. High
simulated intensities in the interval of 1000 to 1500 CU of the
observed intensity in Figure \ref{fig:dust}(d) correspond to this
region. On the other hand, the simulated intensity is lower than the
observed one in the region around (\textit{l}, \textit{b}) =
(6\degr, -28\degr). Low simulated intensities in the interval of
1700 to 2400 CU of the observed intensity in Figure
\ref{fig:dust}(d) correspond to this region. The region is affected
by scattering from the foreground local dust of the photons of a
bright star HD189103 located 190 pc from the Sun. HD189103 is the
brightest UV star in the region of consideration shown in Figure
\ref{fig:dust} according to the TD-1 catalogs. With these three
regions excluded, together with the region of \textit{E(B-V)} less
than 0.07 where dust scattering effect is expected to be small, the
correlation coefficient and the reduced chi-squared value are
estimated to be 0.94 and $\chi^2$ = 3.18, respectively, with 790
degrees of freedom. The rather large value of the reduced chi-square
may indicate that the foreground contamination still remains,
especially in the low latitude region near the Galactic plane.

\begin{figure}
 \begin{center}
  \includegraphics[width=7cm]{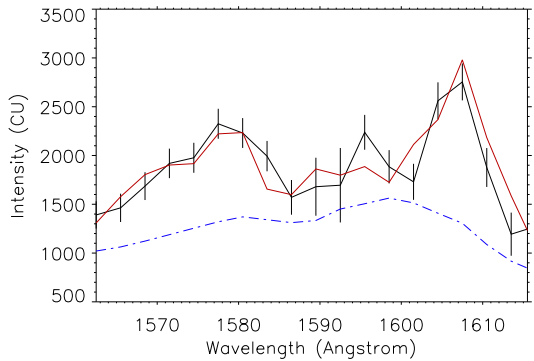}\\
 \end{center}
\caption{Observed spectrum (black solid line) of the lower
supershell with a thickness of $\sim$3\degr, excluding the R CrA
cloud region, and the best-fit result of the PDR simulation (red
solid line). The blue dash-dotted line is a model continuum
emission. \label{fig:pdr}}
\end{figure}

\subsection{Molecular hydrogen fluorescent emission}

Most of the H$_2$ fluorescent emission seen in Figure \ref{fig:spec}
originates from the lower supershell region. We modeled this region
as a PDR with the radiative transfer simulation code {\sc CLOUD}
\citep{van86,bla87}. We took the lower supershell region below
\textit{b} = -15\degr with a thickness of $\sim$3\degr, excluding
the R CrA cloud region, whose spectrum is shown in Figure
\ref{fig:pdr} as a black solid line with the spectral resolution of
the instrument, which is 3 {\AA} around 1600 {\AA}. We fit the H$_2$
fluorescence lines only for the spectral intervals from 1562 to 1616
{\AA} to avoid the effects from the ion emission lines. The
continuum level of FUV emission is shown as a blue dash-dotted line,
which generally follows the minimum data points of the spectrum. We
assume I$_{UV}$, a scaling factor representing the local ISRF
normalized by the model of \citet{dra78}, to be 2, in view of the
ISRF intensity of $\sim$3.5 $\times$ 10$^5$ photons s$^{-1}$
cm$^{-2}$ {\AA}$^{-1}$ given by {\sc GALPROP} at 1300 {\AA}. The
trial values and the ranges of parameters in the simulation are as
follows: the density \textit{n$_H$} = [1, 3, 10, 30, 100] cm$^{-3}$,
the temperature \textit{T} = [10, 30, 100, 300, 1000] K, and the
column density of molecular hydrogen \textit{N(H$_2$)} =
10$^{[15.0-21.0]}$ cm$^{-2}$ with 10$^{0.5}$ steps. All 325 possible
combinations of these fitting parameters were explored in our
simulations.

The best-fit model is shown in Figure \ref{fig:pdr} as a red solid
line with the parameters of \textit{n$_H$} = 30 cm$^{-3}$,
\textit{T} = 300 K, and \textit{N(H$_2$)} = 10$^{18.5}$ cm$^{-2}$.
The reduced chi-squared value is $\chi^2$ $\sim$ 1.36 with 18
degrees of freedom. The 1-sigma confidence ranges are as follows:
\textit{N(H$_2$)} = 10$^{18.0-20.0}$ cm$^{-2}$, \textit{n$_H$}
$\geq$ 10 cm$^{-3}$, while the temperature is not constrained over
the range of 10--1000 K. In fact, the average flux of H$_2$
fluorescent emission was hardly affected over the given range of
other parameters when temperature was increased up to $\sim$5000 K.
As we approximate the total hydrogen column density N$_H$ to be the
neutral atomic hydrogen column density of $\sim$ 4.7 $\times$
10$^{20.0}$ cm$^{-2}$ suggested by \citet{mos12} because the overall
contributions by N(H+) and N(H$_2$) are expected to be minimal, the
range of the molecular hydrogen column density \textit{N(H$_2$)} =
10$^{18.0-20.0}$ cm$^{-2}$, when divided by the total hydrogen
column density N$_H$, yields the H$_2$ abundance, N(H$_2$)/N$_H$, to
be 2 $\times$ 10$^{-3}$ -- 2 $\times$ 10$^{-1}$. This result is
reasonable when compared with the model shown in Figure 31.3 of
\citet{dra11}, in which the H$_2$ abundance increases from below
10$^{-6}$ for the warm neutral medium (WNM) to above 10$^{-1}$ for
the cold neutral medium (CNM), although the transition was obtained
for a slightly different value of equilibrium pressure (p/k = 3000
cm$^{-3}$ K while p/k = 9000 cm$^{-3}$ K in the present simulation)
and occurred at N$_H$ of $\sim$8 $\times$ 10$^{20}$ cm$^{-2}$.

\begin{figure}
 \begin{center}
  \includegraphics[height=8.0cm]{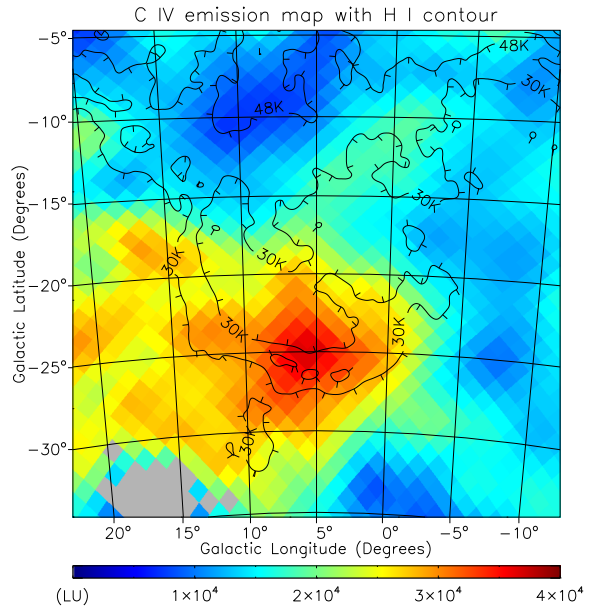}
 \end{center}
\caption{\mbox{C\,{\sc iv}} emission map shown with \mbox{H\,{\sc
i}} contours at the systemic velocity of 7 km s$^{-1}$ overplotted.
The \mbox{C\,{\sc iv}} intensities are given in LU (photons s$^{-1}$
cm$^{-2}$ sr$^{-1}$) The data points in gray color on the bottom
left denote low signal-to-noise ratios of less than 3.
\label{fig:civ}}
\end{figure}

A \mbox{C\,{\sc iv}} emission line is prominently seen in the FUV
spectrum of Figure \ref{fig:spec}. As a \mbox{C\,{\sc iv}} emission
is generally observed in the interface region between hot gas and
cold medium at a transition temperature of T $\sim$ 10$^5$ K, it
should be interesting to locate the origin of this important cooling
line. We have constructed a spectral map for the \mbox{C\,{\sc iv}}
emission line, as shown in Figure \ref{fig:civ}, with its
intensities given in Line Units (photons s$^{-1}$ cm$^{-2}$
sr$^{-1}$, henceforth LU) integrated over the wavelength range of
1540--1560 {\AA}. The map is plotted with a 1\degr pixel size, but
an averaged spectrum of 4\degr centered around each pixel was fitted
to enhance the signal-to-noise ratios (S/N).

We see that, while the \mbox{C\,{\sc iv}} emission peaks in the
bottom region of the shell, the large angular extent of the emission
region across the shell over more than ten degrees, without any
resemblance to the shell-like morphology, indicates that the
emission may not be related to the structure of the present
supershell. Furthermore, the high intensity of 3 $\times$ 10$^4$ LU
may not be consistent with the quite evolved ($\sim$15 Myrs old) and
distant nature of the shell. For example, the peak \mbox{C\,{\sc
iv}} intensity was estimated to be $\sim$6000 LU at the boundary of
the Orion-Eridanus superbubble whose age is a few Myrs, much younger
than GSH 006-15+7. \citet{mos12} also found that soft X-rays do not
show any convincing features associated with the supershell. Hence,
we believe the bright \mbox{C\,{\sc iv}} emission in this region is
of local origin, perhaps the Local Hot Bubble or Loop I superbubble.
In this regard, we note that \citet{sal08} observed \mbox{C\,{\sc
iv}} intensity above 10$^4$ LU toward the interaction zone of the
Local Hot Bubble and Loop I.


\section{Summary}

We have reported here the results of analysis based on the FUV
observations of GSH 006-15+7 made with \textit{GALEX} and
\textit{FIMS}. The main findings are as follows.

1. FUV emission is seen to be enhanced above 2000 CU in the lower
supershell region (\textit{b} $<$ -15\degr) of the supershell
region, which was demonstrated to mostly come from the dust
scattering of interstellar photons of the region.

2.  Monte Carlo simulation for dust scattering was performed for the
lower supershell region under the theoretical albedo of 0.4 and
asymmetry factor of 0.6 and the distance to GSH 006-15+7 was
estimated to be 1300 $\pm$ 800 pc, corresponding well with the
previous estimation of 1500 $\pm$ 500 pc based on kinematics.

3.  Molecular hydrogen fluorescence features are seen in the
spectrum of the lower supershell region: a PDR simulation indicated
an H$_2$ column density of N(H$_2$) = 10$^{18.0-20.0}$ cm$^{-2}$ and
the total hydrogen density of n$_H$ $\sim$ 30 cm$^{-3}$.

4.  An enhanced \mbox{C\,{\sc iv}} emission is seen in the region of
present consideration, but it originates from a broad region across
the shell, indicating its irrelevance to the shell.

\acknowledgments

The authors appreciate the anonymous referee for the valuable
comments which have significantly improved the paper.
\textit{FIMS/SPEAR} was funded by the Ministry of Science and
Technology (Korea) and a grant NAG5-5355 (NASA). It is a joint
project of the Korea Astronomy and Space Science Institute, the
Korea Advanced Institute of Science and Technology, and the
University of California at Berkeley (USA). K.-W. Min acknowledges
the support by the National Research Foundation of Korea through its
Grant No. NRF 2012M1A3A4 A01056418.

\clearpage

\end{document}